\documentclass[conference]{IEEEtran}
\IEEEoverridecommandlockouts

\usepackage{cite}
\usepackage{amsmath,amssymb,amsfonts}
\usepackage{algorithmic}
\usepackage{graphicx}
\usepackage{textcomp}
\usepackage{xcolor}
\usepackage{subfigure}

\usepackage{booktabs}
\usepackage{multirow}
\usepackage{booktabs,caption}
\usepackage[flushleft]{threeparttable}
\usepackage{flushend}

\usepackage{url}  
\def\BibTeX{{\rm B\kern-.05em{\sc i\kern-.025em b}\kern-.08em
    T\kern-.1667em\lower.7ex\hbox{E}\kern-.125emX}}
    
\usepackage{color, colortbl}
\definecolor{Gray}{gray}{0.9}
\usepackage{multirow}

\UseRawInputEncoding
    
\begin{document}

\title{Digital Twin-based Intrusion Detection for Industrial Control Systems}

\author{Seba Anna Varghese$^{1,2}$, Alireza Dehlaghi Ghadim$^{2}$, Ali Balador$^{2}$, Zahra Alimadadi$^{1}$ and Panos Papadimitratos$^{1}$\\
	   $^1$Networked Systems Security Group, KTH Royal Institute of Technology, Stockholm, Sweden\\ \{vargh, alimadad, papadim\}@kth.se\\
       $^2$RISE Research Institute of Sweden, V\"{a}ster\aa s, Sweden, \{alireza.dehlaghi.ghadim, ali.balador\}@ri.se\\
       }

\maketitle

\begin{abstract}
Digital twins have recently gained significant interest in simulation, optimization, and predictive maintenance of Industrial Control Systems (ICS). Recent studies discuss the possibility of using digital twins for intrusion detection in industrial systems. Accordingly, this study contributes to  a digital twin-based security framework for industrial control systems, extending its capabilities for simulation of attacks and defense mechanisms. Four types of process-aware attack scenarios are implemented on a standalone open-source digital twin of an industrial filling plant: command injection, network Denial of Service (DoS), calculated measurement modification, and naive measurement modification. 
A stacked ensemble classifier is proposed as the real-time intrusion detection, based on the offline evaluation of eight supervised machine learning algorithms.
The designed stacked model outperforms previous methods in terms of F1-Score and accuracy, by combining the predictions of various algorithms, while it can detect and classify intrusions in near real-time (0.1 seconds). This study also discusses the practicality and benefits of the proposed digital twin-based security framework.

\end{abstract}


\section{Introduction}

Industrial Control Systems (ICS) are responsible for real-time system monitoring and automatic control of critical industrial infrastructures  \cite{article}, \cite{MLVuln}. ICS use industrial communication protocols, typically lacking built-in security mechanisms, being developed originally for closed environments \cite{article},\cite{csse.2021.014384}. Moreover, with Industry 4.0, systems are increasingly connected to the Internet and, therefore, more exposed to cyberattacks \cite{HAItestbeddataset}. 
Cyberattacks against critical infrastructures, such as cyberattacks on nuclear facilities in Iran 
\cite{Stuxnet}, the Ukrainian power grid 
\cite{Blackenergy}, and natural gas pipeline companies in the US 
\cite{IDSinICS}, 
motivate the importance of effective Intrusion Detection Systems (IDS) for industrial systems.

However, deploying IDS on top of time-critical Programmable Logic Controllers (PLCs) remains a significant challenge: it is crucial to deploy security in ICS that does not affect the smooth running of  tight control loops and operations. 
Another challenge is to implement and maintain security testbeds for IDS. Using operational ICS as a testbed is not allowed due to confidentiality and safety issues. Moreover, having physical security testbeds is quite expensive and time-consuming, usually resulting in incomplete and outdated setups \cite{SecurityAware}. To tackle these problems, digital twin-based IDS and testbeds tend to be practical solution.

A digital twin is a virtual representation of a physical system that can mirror characteristics of its physical counterparts in near real-time \cite{DTdefinition}. The digital twin covers the whole life-cycle of a physical system and represents an up-to-date version of the physical system \cite{Unleash}. Although the key idea behind digital twin is to enhance the manufacturing system life-cycle, some recent works emphasize using digital twins to enhance ICS security. Leveraging a digital twin can facilitate detecting security threats and possibly  sending process control alarms to the physical twin, to take preventive measures. Digital twin-based security analysis does not run on constrained devices such as PLCs and hence offers the possibility to utilize methods that require increased computing resources for cybersecurity analysis (e.g., Machine Learning (ML) and deep learning) \cite{IDSinICS}. Moreover, with digital twins, the security analysis outside the real infrastructure avoid disruptions and damage caused on the actual system \cite{DTsmartgrids}. 

Commercial ICS digital twin solutions are not publicly available for research purposes, as they might make the underlying ICS vulnerable.
Moreover, existing open-source ICS digital twins are not suitable for security research lacking cyber-attack implementations and dataset generation for ML-based IDS development. This is where the contribution of this work lies, contributing to an existing ICS Digital Twin based framework \cite{SIEM}. Given the usefulness of the simulation-based analysis of attacks and counter-measures, it is important to broaden the gamut of attacks and defense mechanisms available on the Digital Twin side. In particular:

\begin{enumerate}
    \item We extend an open-source ICS digital twin framework with an ML-based IDS module.
    \item We implement different types of process-aware attacks in the digital twin, for ICS security monitoring.
    \item Based on the comparison of diverse supervised ML classification algorithms, we design and evaluate an ML-based IDS using a stacked ensemble classifier model.
\end{enumerate}
The remainder of the paper is organized as follows: Section \ref{background} presents the background and related works. A detailed description before we conclude the proposed framework components is provided in Section \ref{mainwork}. Section \ref{experimentalsetup} explains the experimental setup and results are in Section \ref{results}, before we conclude.


\section{background and state of the art}
\label{background}

\subsection{Digital Twin}
\label{labeldigitaltwin}

There are two possible implementations for ICS digital twins: (i) information/knowledge-driven, and (ii) data-driven \cite{AYODEJI20202687}. The former uses physical system specifications to model the virtual system prototypes, while,  the latter uses real-time data from devices in the physical environment as inputs to form a system model. We use the knowledge-driven digital twin approach, along with standalone simulation, without actual physical implementation. Therefore, we discuss related digital twins works in this category for security purposes.

\cite{SecurityAware} proposed a security-aware \textit{CPS Twinning} framework that automatically generates the digital twin of an ICS from its specification. This framework supports two operation modes: (i) simulation mode, with the digital twin run as a standalone simulation, and (ii) replication mode that supports synchronization between the digital twin and physical twin. ~\cite{SecurityAware} focuses on rule-based IDS, while we take an  ML-based IDS approach. 

A digital twin-based security architecture for Industrial Automation and Control Systems (IACS)\cite{DTIACS} focuses on the digital and physical twin synchronization using an active state replication approach, and also lists the security requirements for the proposed architecture; intrusion detection is mentioned briefly and IDS implementation is left for future work. Dietz et al. \cite{SIEM} demonstrated the feasibility of integrating digital twin security simulations in a security operations center, 
using MiniCPS-based \cite{MiniCPS} digital twin and security analytics performed with a SIEM module that uses a rule and log-based incident detection. However, this framework cannot detect process-aware attacks, such as false data injection. \cite{IDSinICS} proposed an IDS in a digital twin environment, with the  IDS comparing the Kalman Filter estimated output to the real system output to detect anomalies; without relying on actual physcial twin or ICS.

\subsection{ML-based IDS}
\label{labelmlbasedids}
\cite{ML3} provided analysis of supervised and unsupervised ML-based anomaly detection in ICS, using datasets of process measurements collected from a Secure Water Treatment testbed.
The results of using supervised algorithms, such as Support Vector Machine (SVM), Random Forest (RF), and K Nearest Neighbor (KNN), is compared to the results of using unsupervised ones; supervised algorithms have better accuracy and F1-score. Furthermore, RF has the best accuracy and F1-score among supervised algorithms. However, the problem considered is a binary classification for one attack category. \cite{MLVuln} evaluates seven supervised ML algorithms for detecting intrusions on a traffic dataset collected from a SCADA water treatment testbed. The algorithms used are SVM, KNN, RF, Naïve Bayes (NB), Decision Tree (DT), Logistic Regression (LR), and Artificial Neural Network (ANN). The RF classifier provides the best results in terms of accuracy, false alarm rate, and undetected rate. One of the future directions mentioned in this paper is a joint design of multiple algorithms to achieve better performance. \cite{ML1} compared SVM, RF, KNN, and One Rule (OneR) algorithms for network intrusion detection using a gas pipeline dataset. It also discussed the advantage of using Particle Swarm Optimization (PSO) and showed that an RF classifier optimized by PSO gives the best results.

\section{Proposed Components}
\label{mainwork}
We first explain the proposed security framework for ICS in Section \ref{securityframework}. Providing a labeled dataset for the evaluation of the designed IDS, multiple attacks scenarios are modeled in Section \ref{attackmodel}. Finally, Section \ref{designids} discusses the ML-based IDS implementation.
\subsection{Security Framework for ICS}
\label{securityframework}
\begin{figure}[!ht]
  \begin{center}
    \includegraphics[scale=0.24]{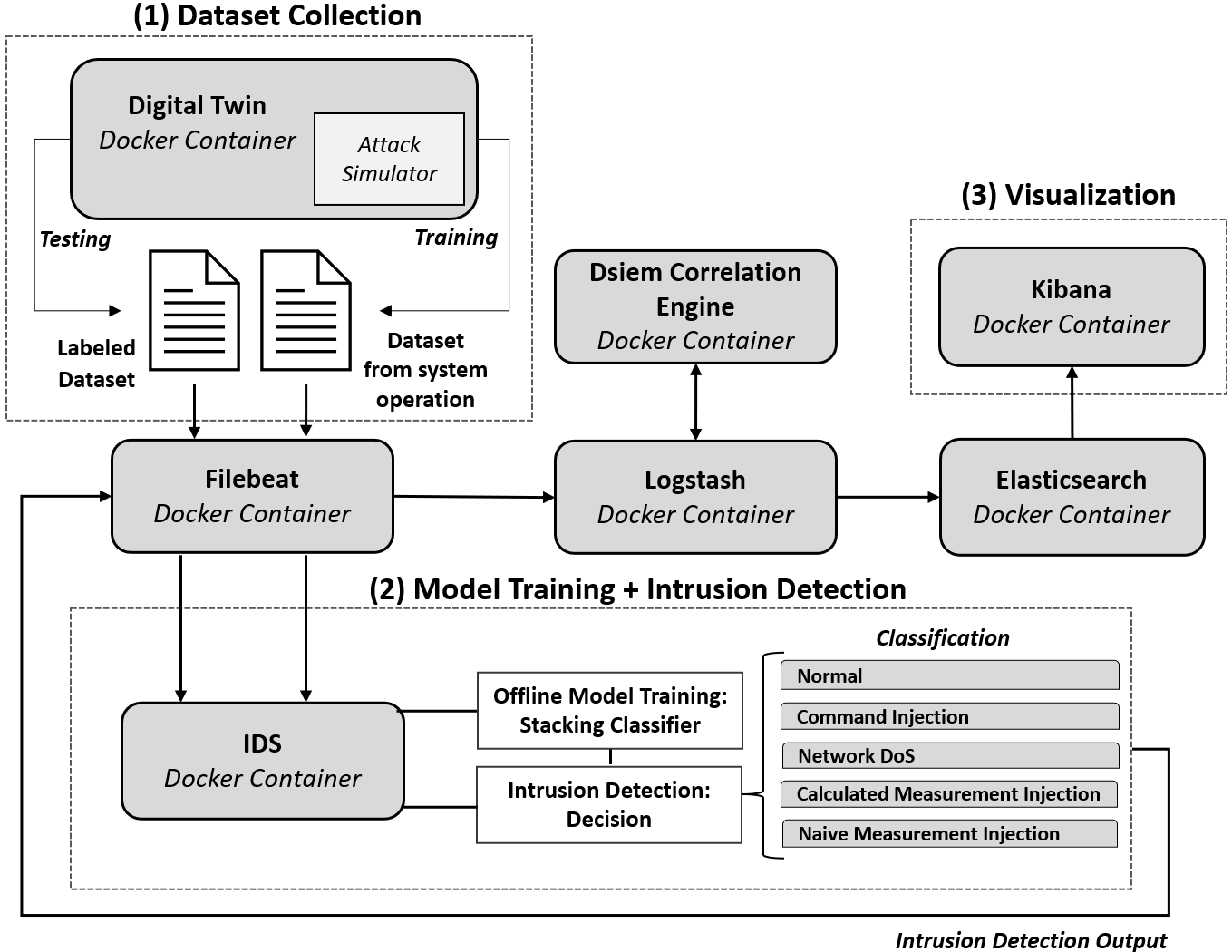}
      \caption{Enhanced Digital Twin security framework, by adding an ML-based IDS and attack simulation modules to the framework proposed in \cite{SIEM}.}
  \label{fig:idsframework}
  \end{center}
\end{figure}

We extended the framework proposed in \cite{SIEM} by adding an ML-based IDS module and implementing various cyberattacks in the digital twin module. Figure \ref{fig:idsframework} depicts the security framework, with all modules implemented as Docker containers. We briefly discuss each module, with more details available in \cite{SIEM}.

\begin{figure}[!ht]
  \begin{center}
    \includegraphics[width=0.45\textwidth]{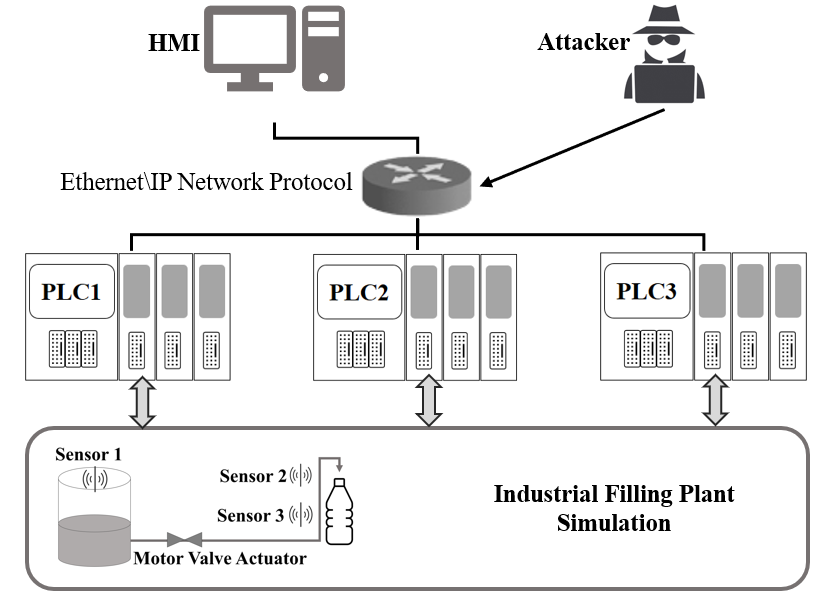}
  \end{center}
  \caption{Industrial filling plant use case \& network topology.}
  \label{fig:dt}
\end{figure}

The security framework in Figure \ref{fig:idsframework} is composed of the following modules:

\begin{itemize}

    \item \emph{\textbf{Digital Twin}} module: runs the MiniCPS-based simulation of an industrial filling plant as a hardware in a loop process. In this, we execute different modeled attacks to simulate malicious activities of intruders and malware.

     \item \emph{\textbf{Filebeat}} module: used to ship the system logs from the digital twin module to the Logstash Module. We extend this to deliver the dataset containing process measurements from the digital twin module to IDS Module.
    
    \item  \emph{\textbf{Logstash}} module: responsible for parsing the log files coming from Filebeat and for producing structured data.
    \item \emph{\textbf{IDS}} module: implemented using an ensemble stacked classifier, classifying data samples from the digital twin.
    \item \emph{\textbf{DSiem Correlation Engine}}: responsible for incident detection based on a rule-based correlation engine that monitors system logs.
    \item \emph{\textbf{Elasticsearch}} module: responsible for data storage. It also receives search queries from the Kibana module and executes them over structured logs. 
    \item \emph{\textbf{Kibana}} module: used for visualizing the results of the SIEM module and the IDS module. We introduce a new dashboard in Kibana to display the the IDS module results.
\end{itemize}
The ICS simulation architecture and network topology is shown in Figure \ref{fig:dt}: consisting of three PLCs, one Human Machine Interface (HMI), an attacker node, and industrial filling process simulation as a physical process. The physical process consists of a liquid tank, a bottle, and a connecting pipe. A motor valve actuator controls the liquid flow from the tank to the bottle. Sensors 1, 2, and 3 read the liquid level in the tank, the flow level in the pipe, and the liquid level in the bottle, respectively. PLC1 monitors and controls sensor 1 and the motor valve actuator; PLC2 and PLC3 are responsible for sensors 2 and 3, respectively. PLC1 performs the control operation of the actuator based on all three sensor measurements. This simulation uses Ethernet/IP (ENIP) as the industrial network communication protocol.

\subsection{Attack Model}
\label{attackmodel}

 Assuming the attacker knows the ICS process and architecture, we model and execute different process-aware attack scenarios as insider threats in the digital twin container. Attacks disturb the physical processes in the real world; here, the process measurements collected from the digital twin reflect the consequence of attacks happening in the physical system. 
Denial of service and data/command injection attacks are significant threats for industrial control systems \cite{MLVuln,faramondi2021hardware}. A few articles take steps further, dividing injection attacks into different scenarios \cite{HAItestbeddataset, SyncFatemah,morris2015industrial}. In this regard, a total of 23 attack scenarios for four different attack types, namely, \emph{command injection}, \emph{network Denial of Service (DoS)}, \emph{calculated measurement modification}, and \emph{naive measurement modification}, are modeled and executed in the digital twin. A summary of the implemented scenarios is shown in Table \ref{tab:scenarios}, while the source code and detailed explanation of the implemented attacks are available in github\footnote{\url{https://github.com/sebavarghese/DT-based-IDS-framework}}.
 
 \begin{itemize}
    \item
\emph{\textbf{Command injection}} exploits the PLC1 interface to remotely inject commands maliciously control the motor valve. This attack exploits the lack of authentication in the ENIP protocol.
\item
\emph{\textbf{Network DoS}} clogs the PLC1 and prevent it from receiving any measurements, or selectively erase specific measurements from reaching PLC1.
\item
\emph{\textbf{Calculated measurement modification}} is a false data injection/modification attack exploiting the lack of encryption in the ENIP protocol. The measurements reaching the PLC1 are altered by a calculated value (positive or negative scaling), they are performed gradually and carefully, aiming to disturb the system operation while avoiding detection.  
\item
\emph{\textbf{Naive measurement modification}}
is similar to the calculated measurement modification, except that the sensor measurements are altered to a constant/random value within the operating range of the system. 

\end{itemize}

\begin{table*}[]
\begin{center}

  \begin{threeparttable}
\centering
\caption{Summary of 23 Attack scenarios implemented on digital twin module.}
\label{tab:scenarios}

\begin{tabular}{@{}llll@{}}
\toprule
Attack Type                  & Num & Attack Description                             & Target          \\ \midrule \midrule

Command Injection            & 1   & Toggle actuator value every 0.5 Second         & MS              \\ \\
\multirow{2}{*}{Network DoS} & 3   & Drop Packets (MitM attack)                    & FL, BLL, FL+BLL \\
                             & 1   & TCP SYN flood attack targeting ENIP port 44818 & FL+BLL          \\ \\
\begin{tabular}[c]{@{}l@{}}Naive Measurement Modification\end{tabular} &
  6 &
  \begin{tabular}[c]{@{}l@{}}Change the value to a constant value\\ Change the value to a uniform random variable\\  \end{tabular} &
  \begin{tabular}[c]{@{}l@{}}FL, \\ BLL, FL+BLL\end{tabular} \\
\begin{tabular}[c]{@{}l@{}}Calculated Measurement Modification\end{tabular} &
  12 &
  \begin{tabular}[c]{@{}l@{}}\\Scale down/up by random value\\ Scale down/up by fixed value\\ \\\end{tabular} &
  \begin{tabular}[c]{@{}l@{}}FL, \\ BLL, FL+BLL\end{tabular} \\ \bottomrule
\end{tabular}%

    \begin{tablenotes}
      \small
      \item \textbf{MS:} Motor State on PLC1, \textbf{FL:} Flow Level on PLC2, \textbf{BLL:} Bottle Liquid Level on PLC3.
    \end{tablenotes}
  \end{threeparttable}
\end{center}
\end{table*}

\subsection{ML-based IDS}
\label{designids}
We use off-the-shelf classification techniques. We identified eight supervised ML-based IDS techniques relevant to ICS, namely SVM, KNN, NB, RF, LR, ANN, DT, and Gradient Boosting (GB). 
We created a labeled dataset, by logging the system state in the presence of each process-aware attack scenario as well as the attacker-free behavior of the system. The process measurements collected during attacker-free operation are labeled as 'Normal' and those during attacks are labeled with the corresponding attack category. The dataset is split into two subsets: 70\% of the dataset for training, and the remaining 30\% for testing the models. We evaluated the algorithms based on typical performance parameters such as, \emph{confusion matrix}, \emph{accuracy}, \emph{precision}, \emph{recall}, and \emph{F1-score}.

Moreover, we use an ensemble approach, \emph{stacking}, to design a signature-based IDS. This IDS combines individual classifiers and makes the final inference as the most closely corresponding class. The individual classifiers used in this work are the ones that showed relatively the best performance in previous studies, as mentioned in Section \ref{labelmlbasedids}. Therefore, an ensemble classifier using these individual classifiers can achieve good performance.

We use two levels: Level 0 and Level 1. 
Level 0 has three individual classifiers, and Level 1 is the final classifier. The choice of Level 0 classifiers is based on the evaluation results of the eight individual classifiers on distinct class labels. Typically, the classifier that gives the best overall scores across all class labels may not be the best one for each label. Predictions from Level 0 classifiers are represented by P1, P2, and P3. The Level 1 classifier used is a neural network using a Multi-Layer Perceptron (MLP) classifier that combines the outputs of Level 0 classifiers. Here, the Level 1 classifier is trained using the cross-validated predictions from Level 0 classifiers \cite{StackingMechanics}. Finally, the stacked model is chosen as the classifier model to implement the IDS.

\begin{figure}[!ht]
  \begin{center}
    \includegraphics[width=0.48\textwidth]{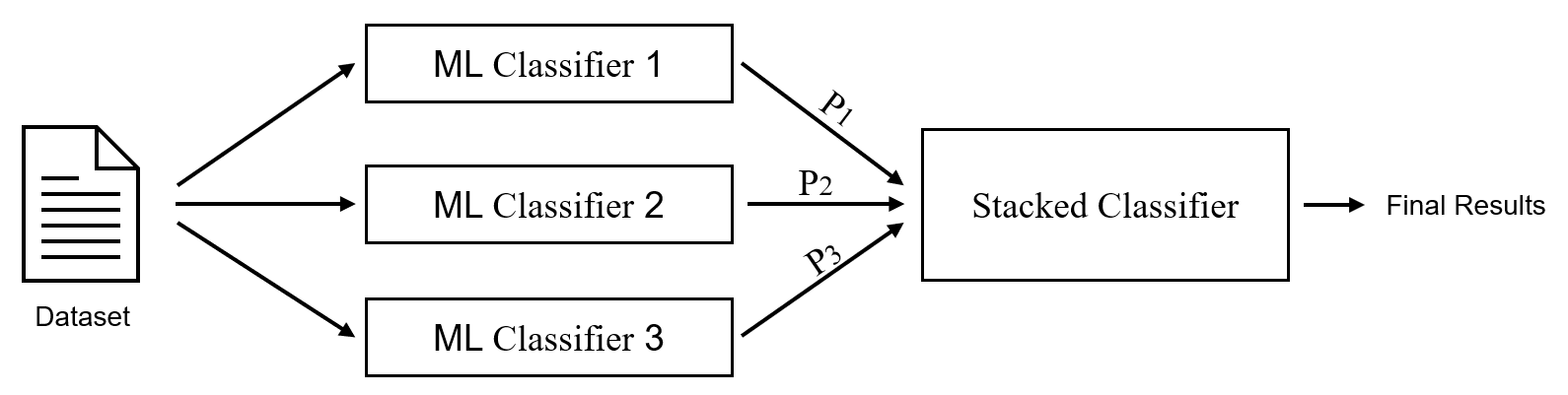}
  \end{center}
  \caption{Stacked ensemble classifier model concept.}
  \label{fig:Stacking}
\end{figure}

\section{experimental setup}
\label{experimentalsetup}

\textbf{Hardware and Technical Stack:}
\label{technicalstack}
The framework runs on a Linux (Ubuntu 18.04) Virtual Machine hosted on a Microsoft Windows 10 machine.  The detailed system setup and technical specifications to run the simulation are presented in Table \ref{tab:systemspec}. We add the IDS module as a Docker container to the initial framework. We implement and evaluate the supervised ML algorithms using the \emph{Scikit-learn}\footnote{\url{https://scikit-learn.org/stable/}} library for Python.

\begin{table}[!h]
\centering
\caption{System used to run ML algorithms}
\label{tab:systemspec}
\begin{tabular}{c c} 
 \hline
 \textit{Parameters} & \textit{Values} \\
 \hline\hline
 \rowcolor{Gray}
\textbf{System Type} & x64-based PC \\
\textbf{OS Name} & Microsoft Windows 10 Home \\
\rowcolor{Gray}
\textbf{OS Version} & 10.0.19043 Build 19043 \\
\textbf{Processor} & \begin{tabular}[c]{@{}c@{}}Intel(R) Core(TM) i5-6400 CPU @ 2.70GHz,\\  4 Core(s) , 4 Logical Processor(s)\end{tabular} \\
\rowcolor{Gray}
\textbf{System RAM} & 16GB \\
\textbf{VM RAM} & 4GB \\
 \hline

\end{tabular}
\end{table}

\textbf{Attack Implementation:}
\label{attackimplementation}
We discuss how different process-aware attacks are modeled and executed in the digital twin module.
\begin{enumerate}
\item

\emph{\textbf{Command injection:}} 
Exploiting the lack of authentication vulnerability, we read the actuator value from PLC1 and forge the commands to toggle the valve state on PLC1, using a custom Python script.
\item
\emph{\textbf{Network DoS:}} The first approach is to clog the network using TCP SYN flooding on TCP port 44818 (ENIP communication port) of PLC1. The second one selectivly erase messages reaching PLC1 using a combination of Man in the Middle (MitM) and ARP poisoning using Ettercap\footnote{\url{https://www.ettercap-project.org/}} tool. As an illustration, in one scenario, the attacker node placed in between PLC1 and PLC2 sniffs the packets sent towards PLC1 from PLC2 using ARP poisoning, and drops these packets. In this case, the attacker selectively erases the packets reaching PLC1 from PLC2, where PLC1 can still receive packets sent from PLC3. 

\item
\emph{\textbf{Calculated/Naive measurement modification:}} We use custom Python scripts with \emph{scapy}\footnote{\url{https://scapy.net/}} to decode and alter packets sent on the network. 
The attacker alters the measurements sent to PLC1 by launching a MitM attack. For naive measurement modification attacks, these measurements are altered to either a constant or a random value within the predefined limits of the process measurements. 
For calculated measurement modification attack, the measurements are altered to a scaled value using a small factor in the range (0,1], attempting to make the attack stealthy:
\end{enumerate}

\begin{equation}
\resizebox{8cm}{!}{%
$modified\_value = (1 \pm scaling\_factor)\times sensor\_value$
}
\end{equation}

\textbf{Dataset Generation:}
\label{generateddataset}
We collected the process state variables during the attacker-free operation and during the attack execution over 3 hours of the digital twin operation. Meanwhile, we considered sufficient recovery time between attacks to avoid having simultaneous impacts of two attacks at the same time. PLC1 node recorded process state variables every 0.5 seconds and generated a CSV file to serve as a training dataset for the ML-based IDS. The generated data set consists of 2705 records composed of 1920 attacker-free  and 785 anomalous samples. Anomalous samples consist of 434 calculated measurement modification attack samples, 227 naive measurement modification attack samples, 88 network DoS attack samples, and 36 command injection attack samples.
We did not consider balancing classes across normal and under attack samples in the generated dataset to retain the distribution of samples as a real scenario and avoid bias toward fake results. Using unbalanced data is used in \cite{ML2} and \cite{MLVuln}, which emphasize considering the choice of proper evaluation metrics to classify imbalanced datasets instead of using data sampling methods to circumvent the class imbalance.

\section{evaluation results}
\label{results}

\begin{figure}[t]
   \centering
   \subfigure[SVM]{\label{fig:cfm_svm}
     \includegraphics[width=0.4\linewidth]{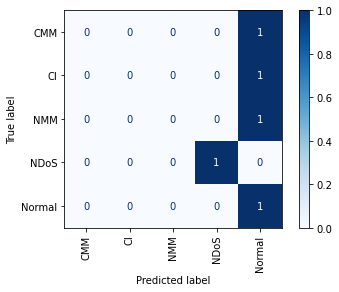}}
   \subfigure[RF]{\label{fig:cfm_rf}
     \includegraphics[width=0.4\linewidth]{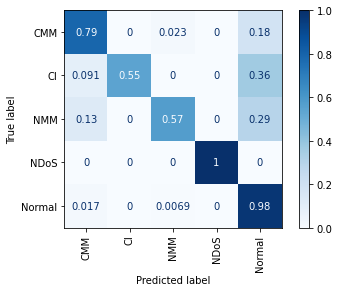}}
   \subfigure[KNN]{\label{fig:cfm_knn}
     \includegraphics[width=0.4\linewidth]{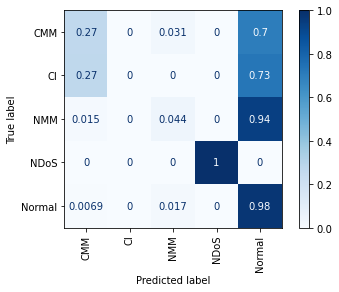}}
   \subfigure[LR]{\label{fig:cfm_lr}
     \includegraphics[width=0.4\linewidth]{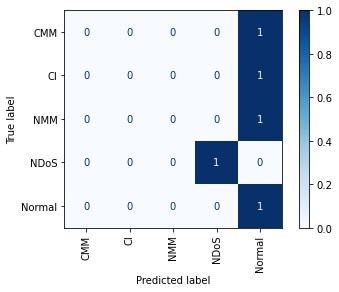}}
   \subfigure[DT]{\label{fig:cfm_dtc}
     \includegraphics[width=0.4\linewidth]{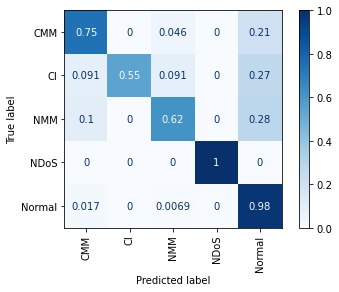}}
   \subfigure[NB]{\label{fig:cfm_nb}
     \includegraphics[width=0.4\linewidth]{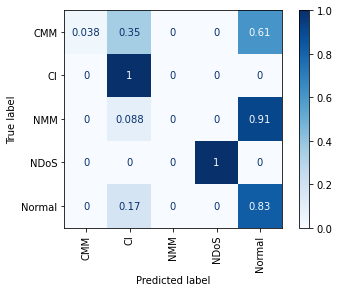}}
   \subfigure[ANN]{\label{fig:cfm_mlp}
     \includegraphics[width=0.4\linewidth]{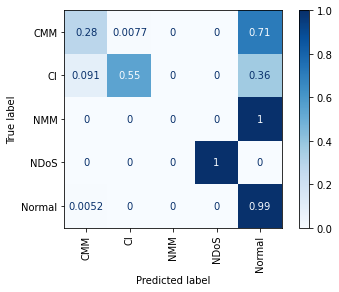}}
   \subfigure[GB]{\label{fig:cfm_gb}
     \includegraphics[width=0.4\linewidth]{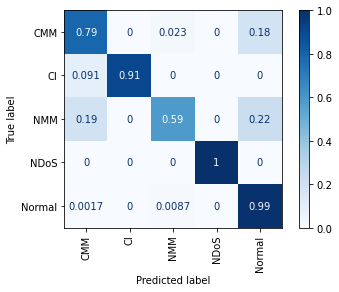}}
   \caption{Normalized confusion matrices of single classifier; for different attacks (Calculated Measurement Modification(CMM), Command Injection (CI), Naive Measurement Modification (NMM), Network DoS (NDoS)).}
      \label{fig:cfmevaln}
\end{figure}

Figure \ref{fig:cfmevaln} shows the normalized confusion matrices for the eight supervised ML algorithms evaluated on our test dataset. In a normalized confusion matrix, the closer to 1 the diagonal elements are, the better the algorithm identifies the corresponding class.

All models correctly identify network DoS attacks; however, their performance for the rest of the attacks: GB is the best classifier for detecting calculated measurement modification attacks and normal; meanwhile DT outperforms the GB model for naive measurement modification attacks, and NB gives the best score for command injection attacks. Therefore, GB, DT, and NB models are chosen as the Level 0 algorithms for the stacked ensemble model.

\begin{figure}[!h]
  \begin{center}
    \includegraphics[width=0.23\textwidth]{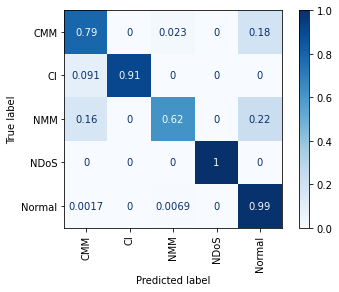}
  \end{center}
  \caption{Normalized confusion matrix for the stacked ensemble based IDS, for CMM, CI, NMM, and NDoS attacks.}
  \label{fig:cfm_stacking}
\end{figure}

\begin{table}
\centering
\caption{Classification performance metrics.}
\label{tab:metrics}
 \resizebox{\columnwidth}{!}{%
\begin{tabular}{lllll}
\hline
\textbf{Algorithm\textbackslash{}Classification Method} &
  \textbf{Accuracy} &
  \textbf{Precision} &
  \textbf{Recall} &
  \textbf{F1-score} \\ \hline
  \hline
 \rowcolor{Gray}
SVM & 0.743 & 0.347                         & 0.4 & 0.37 \\ 
RF  & 0.908                        & 0.921                        & 0.777                       & 0.83                       \\ 
\rowcolor{Gray}
KNN & 0.772                         & 0.553 & 0.458                         & 0.468                       \\ 
LR  & 0.743 & 0.347                        & 0.4 & 0.37 \\ 
\rowcolor{Gray}
DT & 0.904                      & 0.911                         & 0.777                         & 0.828                       \\ 
NB  & 0.64 & 0.568                        & 0.573 & 0.4 \\ 
\rowcolor{Gray}
ANN & 0.792                     & 0.707                         & 0.565                        & 0.594                      \\ 
GB &
  0.924 &
  0.928 &
0.856 &
0.887\\ 
  \rowcolor{Gray}
  Stacked model &
  \textbf{0.927} &
  \textbf{0.936} &
\textbf{0.862}&
  \textbf{0.894}\\ \hline
\end{tabular}
 }
\end{table}

\begin{figure*}[!ht]
  \begin{center}
    \includegraphics[scale=0.47]{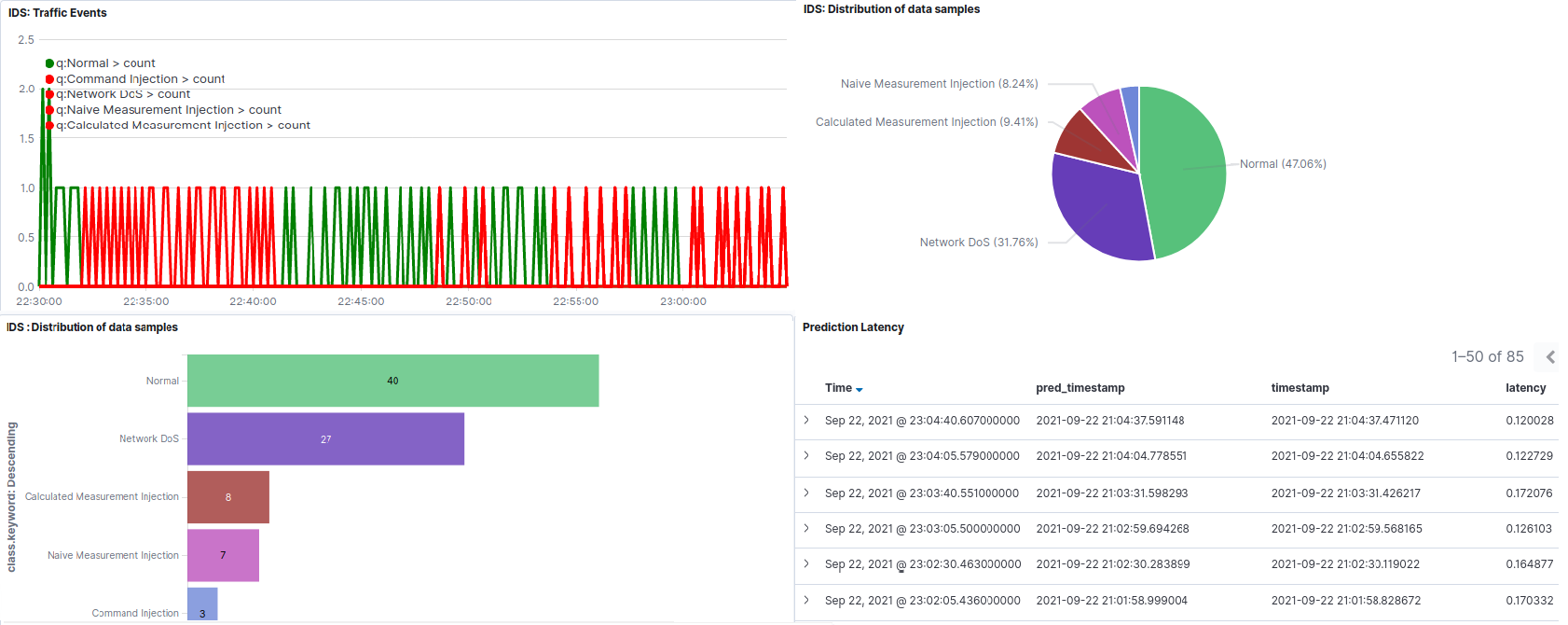}
  \end{center}
  \caption{Screenshot of IDS dashboard.}
  \label{fig:ids}
\end{figure*}

\begin{figure*}[!ht]
    \centering
    \includegraphics[scale=0.47]{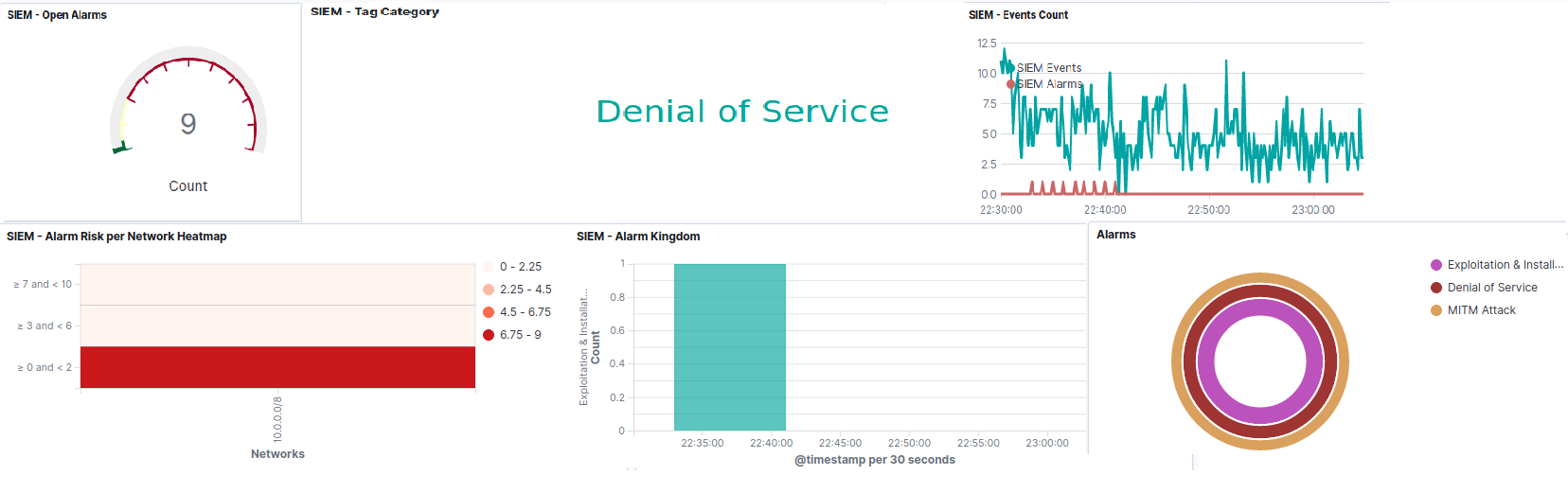}
  \caption{Screenshot of SIEM dashboard.}
  \label{fig:siem}
\end{figure*}

 Figure \ref{fig:cfm_stacking} presents the normalized confusion matrix for the stacked model. The stacked model detects all Network DoS attacks correctly, while it shows scores of 0.99 and 0.91 respectively for normal data and command injection attacks. It also shows classification results for naive measurement modification in the GB model improved respectively from 0.59 to 0.62 with the stacked model. The calculated measurement modification score is 0.79, which improves individual classifiers. Furthermore, Table \ref{tab:metrics} lists the scores of different classification metrics used to evaluate the ML algorithms on the labeled dataset. It shows that the stacked model outperforms the individual classifiers across all the metrics used for evaluation.

The framework proposed in \cite{SIEM} has a module running rule-based correlation engine that checks the severity of the system logs to report attacks. Such a module can only detect those attacks that tamper with the system logs. Figure \ref{fig:siem} is the screenshot of the SIEM dashboard \cite{SIEM}, for a 30-minute period, displaying the results of the correlation engine that uses system logs to identify incidents. The SIEM module does not report any alarm for the executed process-aware attacks, except for network DoS.

To detect other types of attacks like process-aware attacks, we extended framework proposed in \cite{SIEM} with an ML-based IDS module. 
Figure \ref{fig:ids} shows a screenshot of the IDS dashboard in Kibana for the same time frame as that of the IDS dashboard shown in Figure \ref{fig:siem}, with four visualizations: \textit{Traffic Events}, is a time-series visualization of data samples using Timelion\footnote{\url{https://www.elastic.co/guide/en/kibana/current/timelion.html}}. The X-axis represents the timestamps of incoming data samples, whereas the Y-axis represents the number of samples at a given point in time. All data samples classified as 'normal' by the IDS are shown in green, while all those classified as 'attacks' are shown in red. The \textit{pie-chart} visualizes the distribution of data samples across different class labels, in percentage, as classified by the IDS. The bar chart visualizes the \emph{class} labels that IDS identifies for the incoming data samples. The tabular \textit{Latency} visualization provides the latency in classifying the data samples. Here, the columns 'timestamp' and 'pred\_timestamp' indicate the sample collection and prediction time, respectively. The difference between 'pred\_timestamp' and 'timestamp' shows the time it takes to classify.  For the time frame shown in this screenshot, the average latency is 0.1 seconds, which is near real-time.  
Comparing the results in both dashboards shows that ML-based IDS extends the framework to detect a wider range of attacks.

\section{conclusions and future works}
\label{conclusions}

We delivered extensions for a Digital Twin-based  ICS security framework implemented using open-source tools, including an ML-based IDS for detecting intrusions in near real-time. The digital twin is equipped with various process-aware attack scenarios to provide a platform for analyzing and developing intrusion detection/prevention systems. We applied several common ML algorithms to develop an IDS for the Framework. We also designed a stacked model classifier, which improves the IDS classification performance over individual ML algorithms, and can detect cyberattacks in near real-time constrain.

As future work, hyper-parameter tuning of ML algorithms can improve classification scores. Another improvement is using time series-based algorithms to learn the correlation between process measurements across data samples and their changes over time. We can also feed network traffic data to IDS to detect a broader range of attack types, such as reconnaissance attacks. Moreover, evaluation of unsupervised and semi-supervised learning algorithms to detect intrusions is another future work. Such approaches can help detect zero-day attacks and avoid the need for labeling the dataset.

\section*{Acknowledgment}

This work was supported by InSecTT (www.insectt.eu) and DAIS (www.dais-project.eu), which received funding from the KDT Joint Undertaking (JU) under grant agreement No 876038 and No 101007273. The JU receives support from the European Union’s Horizon 2020 research and innovation programme and Austria, Sweden, Spain, Italy, France, Portugal, Ireland, Finland, Slovenia, Poland, Netherlands, Turkey, Belgium, Germany, Czech Republic, Denmark, Norway.
 
The document reflects only the authors’ views and the Commission is not responsible for any use that may be made of the information it contains.

\bibliographystyle{ieeetr}
\bibliography{conference.bib}

\end{document}